# Using the System Schema Representational Tool to Promote Student Understanding of Newton's Third Law

Brant E. Hinrichs

*Department of Physics, Drury University, 900 N. Benton Ave., Springfield, MO 65802*

**Abstract.** The Modeling Instruction program at Arizona State University has developed a representational tool, called a system schema, to help students make a first level of abstraction of an actual physical situation [1]. A system schema consists of identifying and labeling all objects of interest from a given physical situation, as well as all the different types of interactions between the objects. Given all the relevant objects and their interactions, students can explicitly identify which are part of their system and which are not, and then go on to model the interactions affecting their choice of system as either (i) mechanisms for energy transfer, or (ii) forces being exerted. In this paper, I describe the system schema tool, give examples of its use in the context of forces, and present some evidence on its effectiveness in helping students understand Newton's Third Law.

## INTRODUCTION

A strength of the Modeling Instruction (MI) curriculum developed by Arizona State University (ASU) is that it provides students with some basic representational tools for modeling physical objects and processes [1]. Quality tools are seen as vital for helping students build quality scientific models, a central goal of MI [2]. In addition, students who are comfortable using a range of tools in problem solving better approximate physics experts, who routinely create many different representations (e.g. mathematical, graphical, diagrammatic) in the analysis of a single problem [3,4].

Examples of representational tools include: graphs, motion maps, vector diagrams, system schema, pie charts, bar charts, free-body diagrams, etc. [3]. This paper focuses on a description, implementation, and evaluation of just one, the system schema. The second section of the paper describes system schema and gives several illustrative examples. The third section discusses how system schema were implemented in a calculus-based introductory physics course taken by all science majors at Drury University. The last section presents and discusses data from a portion of the Force Concept Inventory (FCI) [5,6], as evidence for the effectiveness of the system schema in helping students understand Newton's Third Law. See also [7] (and especially Jiménez 1999 and 2001 in that paper) for other recent related work in this area.

## WHAT IS A SYSTEM SCHEMA?

A system schema is used to represent an actual physical situation and is the first level of abstraction after a pictorial representation. It serves as a conceptual bridge for students to more abstract representations like free-body diagrams and Newton's Laws. The Physics Teacher has a recent article that details typical implementation of system schema in high school modeling instruction [8]. This paper will instead focus on how system schema have been used in university level introductory courses [9]. There are differences between the two approaches (see below) but it is not the purpose of this paper to highlight them.

Figs. 1 and 2 show typical introductory university level physical situations, corresponding system schema, and free-body diagrams for specified systems within each schema. Figs. 1 and 2 show a static and dynamic situation, respectively.

Although it is not an exhaustive list, here are many of the basic rules that guide students in constructing system schema from a verbal or pictorial representation. (1) All objects are represented without any details of their shape or structure. (2) Two objects interact if they influence one another [10]. For example, in Fig. 1, the book influences the brick (it holds it up) and the brick influences the book (it squashes it a little). (3) Therefore, interactions between objects are represented by two-headed arrows, and labeled with the type of interaction. If object X influences object Y, object Y also influences object X.

This is a built in mechanism that later helps students "explain" Newton's 3rd Law. (4) The introductory level course only deals with four interactions, namely three types of non-contact (gravitational "g", electric "e", and magnetic "m") and one type of contact ("c"). All contact interactions (i.e. normal, frictional, etc.) are lumped into one type, as compared to reference [8] which separates them out. A single type of contact is preferred because it simplifies the process of constructing system schemas, simplifies using them to help construct free-body diagrams (see (10) below), and simplifies the physics model of what an interaction is. (5) Double-headed interaction arrows are drawn as solid lines if the interaction persists for the entire time interval the system schema is of interest; otherwise they are drawn as dashed lines (See Fig. 2b). (6) The gravitational interaction only occurs between a huge object, like the earth or moon, and any other object. (7) A system boundary is represented by drawing a dashed line around one or more objects.

If forces are to be studied, here are rules students use to construct free-body diagrams from a completed system schema. (8) Each object or set of objects for which Newton's Laws are to be written must be identified as a separate system [10] (see Fig. 1). (9) Define Force as a description of an interaction between two objects. (10) For each interaction that crosses a system boundary, there should be one and only one force on the corresponding free-body diagram for that system. Students find this rule enormously helpful especially for complicated systems involving multiple objects with multiple interactions. (11) The label for a force vector in a free-body diagram (e.g. $\vec{F}^c_{B \to R}$) is constructed using the relevant labels from the system schema. See the caption of Fig. 1c for an illustrative example.

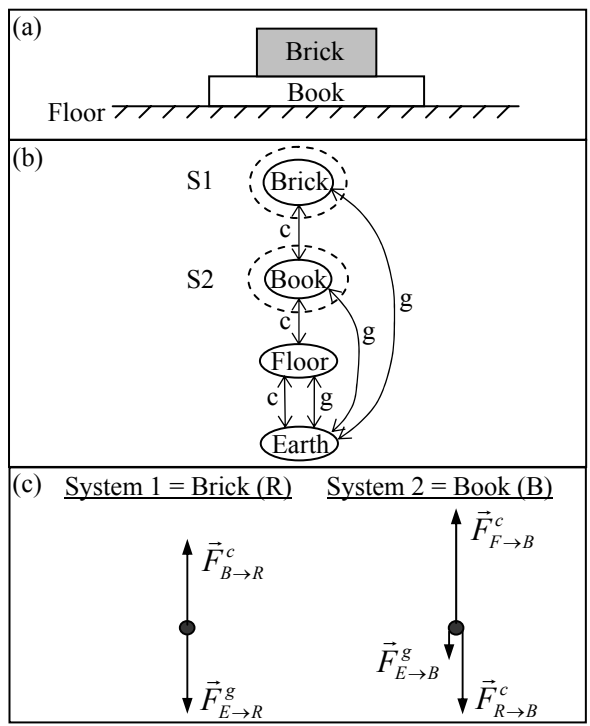

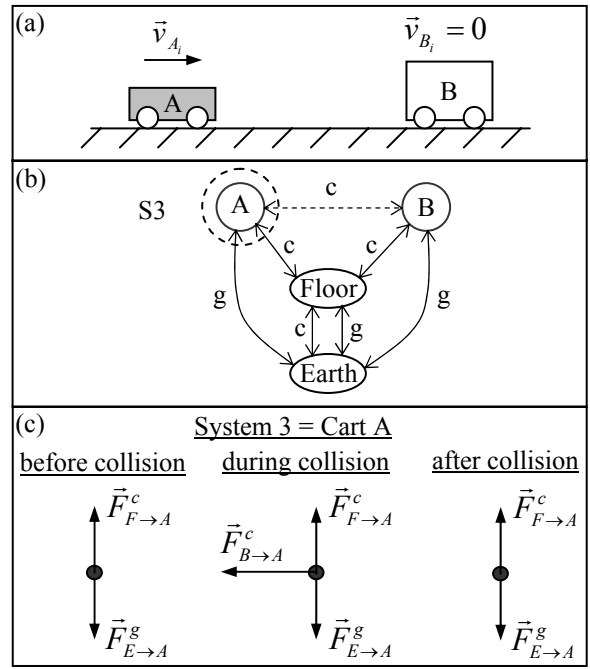

**FIGURE 1.** (a) Pictorial representation of a physical situation. All objects are at rest. (b) System schema of this physical situation, with two of many possible systems identified [10]. The dashed ellipses represent system 1 (S1) and system 2 (S2) respectively. "c" labels a contact interaction, and "g" labels a gravitational interaction. (c) Free-body diagrams for the two systems identified in (b). In a force label "c" means contact, "g" means gravitational. Also, for this particular scenario "B" means book, "R" means brick, "E" means earth, and "F" means floor. For example, the symbol $\vec{F}^c_{B \to R}$ means a contact force by the book on the brick. Note the mass of the brick has arbitrarily been chosen to be three times the mass of the book, thus $\vec{F}^g_{E \to R}$ is three times the length of $\vec{F}^g_{E \to B}$.

**FIGURE 2.** (a) Pictorial representation of an experiment with two low friction carts of differing masses. Cart A is launched toward cart B, which is initially at rest. After the collision (not shown) cart A has reversed direction (not shown). (b) A single system schema for this complete experiment, with one of many possible systems identified. The dashed circle represents system 3 (S3). Note that in contrast to Fig. 1b, this system schema is for a physical situation that changes over time. Before, during, and after the collision is represented here by a single system schema. Therefore, the contact interaction between cart A and cart B is drawn as a dotted line rather than a solid line to indicate that it is not present for the entire time interval of interest of the schema. (c) Free-body diagrams for System 3 at different instants in time in the experiment. Force labels follow the same conventions described in Fig. 1c.

**TABLE 1.** Individual class performance on the Force Concept Inventory (FCI) with and without use of system schema.

| System Schema | Fall Semester | Number of Students (N)[1] | Matched Pre-test Avgs FCI (29)[2] | Matched Pre-test Avgs FCI 3$^{rd}$ Law (4)[3] | Matched Post-test Avgs FCI 3$^{rd}$ Law (4)[3] |
|---|---|---|---|---|---|
| no | 1999 | 17 | 11.1 ± 3.7 | 1.2 ± 1.1 | 2.8 ± 1.2 |
| no | 2000 | 14 | 12.1 ± 5.6 | 1.1 ± 1.0 | 2.7 ± 1.2 |
| transition | 2001 | 9 | 7.3 ± 3.1 | 0.9 ± 0.8 | 2.3 ± 1.4 |
| yes | 2002 | 15 | 10.2 ± 2.7 | 1.1 ± 1.0 | 3.6 ± 1.1 |
| yes | 2003 | 13 | 11.3 ± 3.3 | 1.1 ± 1.0 | 3.8 ± 1.0 |

## CLASSROOM IMPLEMENTATION

In 1999, I began teaching at Drury University and used the Workshop Physics (WP) curriculum [11,12] for the first time. Because I was disappointed in my student's understanding of the 3$^{rd}$ Law, I worked especially hard in 2000 to help them improve, but their scores remained flat (see Table 1, years 1999, 2000).

At a summer 2001 workshop at ASU I was introduced to modeling instruction for the university-level introductory course [9], discourse management [13], and system schema. Fall 2001 I switched from WP to using those methods. As it was a transition period in my teaching, FCI data from that year are not included in any subsequent composite analysis.

The last two years (2002-03) I used the following procedure in teaching Newton's Third Law. Shown the physical situation of Fig. 1a, students groups were asked to predict how the contact force by a 10kg (lead) brick on a 3kg book compared to the force by the book on the brick. They defended their answer with a system schema. Occasionally one group but frequently none came up with the "physics" answer. Most predicted the brick would exert a greater force because it was so much heavier. An instructor demo using force plates to measure the two forces revealed that they are in fact equal in magnitude but opposite in direction. At first, students were puzzled, but then at least one group always noticed that "this makes sense since $\vec{F}_{R \to B}^c$ and $\vec{F}_{B \to R}^c$ are the same interaction." The rest of the class quickly agreed and most seemed satisfied. Here then is the power of the system schema. It gives a visual and conceptual representation, in terms of objects and interactions, for the experimental results showing the 3$^{rd}$ law.

In a later unit on collisions, student groups were shown physical situations similar to Fig. 2a and asked to predict how the force by cart A on B compared to the force by cart B on A. They were asked to make predictions for a range of scenarios, varying both the masses and initial velocities of the two carts. Usually a quarter to a third of groups predicted the "physics" answer. Often these were groups that used a system schema to guide their thinking. Groups then performed experiments using a track, low friction carts, force probes, and a computer interface. Students who predicted incorrectly were initially puzzled by data indicating equal but opposite forces. But when asked whether or not their data was consistent with a system schema, they would draw one, and a light bulb would go off: "System schema even work for collisions! They've got to be the same size because they're the same interaction." Here is, I claim, a conceptual understanding reinforced by a representational tool. Lastly, though most students seemed to understand their data, to reinforce their "aha" moments, the instructor led a short class discussion on Elby's idea of refining our intuition about "reacts more" in a collision to correspond with change in velocity rather than with force [14].

## RESULTS AND DISCUSSION

Student understanding of Newton's Third Law was measured using the Force Concept Inventory (FCI) [5,6]. The FCI was given as a pre-test the first day of class, and then as a post-test at the end of mechanics on the final exam, or on an exam sometime in the second semester during electricity & magnetism. Only data from students who took both the pre-test and post-test are reported here (i.e. the data is matched). All data is from the same instructor, and from the calculus-based introductory physics course taken every year by all science majors at Drury University.

Table 2 shows two composite sets of data. The first row is for classes from 1999 and 2000 that used Workshop Physics (WP) (N=31). The second row is for classes from 2002 and 2003 that used modeling, system schema, and discourse management (N=28).

The pre-test data for both the entire FCI, as well as the four 3$^{rd}$ Law questions, show that students in both populations were statistically similar at the start of the semester. Post-test data show that students in the second population performed much better than the first on the four 3$^{rd}$ Law questions. Levene's Test for Equality of Variances indicated that the variances of the post-test data were different [15], so a 1-tailed, unequal-variance student's t-test was run in Excel. It indicated that we may safely reject the null-hypothesis that the difference in the means is due to random chance (p=.0003 at the 95% confidence level). That is, the difference *is* statistically significant.

---
[1] Class size each year was 19, 14, 18, 18, and 19 respectively.
[2] For historical reasons, Drury uses the original 29 item FCI found in the appendix of [5].
[3] Four FCI Questions - 2, 11, 13, and 14 - test Newton's Third Law.

**TABLE 2.** Composite student performance on the FCI, with and without use of system schema in instruction

| System Schema | Fall Semester | Number of Students (N) | Matched Pre-test Avgs | | Matched Post-test Avgs |
|---|---|---|---|---|---|
| | | | FCI (29)[1] | FCI 3$^{rd}$ Law (4)[2] | FCI 3$^{rd}$ Law (4)[2] |
| no | 1999, 2000 | 31 | 11.6 ± 4.6 | 1.2 ± 1.0 | 2.8 ± 1.2 |
| yes | 2002, 2003 | 28 | 10.7 ± 3.0 | 1.1 ± 1.0 | 3.7 ± 0.8 |

Table 2 shows that modeling, discourse management, and system schema (MDMSS) can lead to excellent scores on the FCI 3$^{rd}$ Law questions. It's possible that students did better in 2002-03 than in 1999-2000 because I was less experienced and confident in my teaching my first years at Drury. However, I was unable to make any improvements in two years of using WP, but I was able to with MDMSS (see Table 1, years 2001 and 2002). Perhaps I just "get" MDMSS better than I "get" WP. Indeed, some recent unpublished data suggest that more experienced WP instructors have data whose means are midway between the two reported here [16].

Although Table 2 does not show if system schema were the primary reason that students did better, a case can be made that it in particular helps students learn Newton's 3rd Law. This is because WP and MDMSS are very similar except for the emphasis in MDMSS on use of representational tools, and especially system schema. Both are similar in their approach to the classroom, emphasizing students working in groups of 3-4, real-time data collection to build models, and a 3-cycle pedagogy of predict, experiment, and resolve-differences. And both are similar in their approach to teaching the 3$^{rd}$ Law as well. Activities based on Fig. 2 in MDMSS are almost identical in style and substance to those in WP. A small difference is that MDMSS also includes the activity based on Fig. 1.

A main difference between MDMSS and WP is an emphasis on system schema. Students start in the third week of the class using system schema to qualitatively model energy concepts [3]. Long before the concept of force has been introduced, students get comfortable talking about objects and interactions. Further, I suggest that spontaneous comments by students about how 3$^{rd}$ Law force pairs are the same size because they are "from the same interaction" indicate something more is going on than just appeal to experimental results. They suddenly seem more satisfied with their experimental results, perhaps because there is a "reasonable" mechanism to "explain" them. Perhaps by the time forces are studied, the system schema has become a valuable epistemological resource to them [14]. It would be an interesting experiment, now that I am more experienced, to return to teaching with WP and see how well my students did with the 3$^{rd}$ Law. And then try teaching with WP but including extensive use of system schema as well.

## ACKNOWLEDGMENTS

Kathy Coy, Kerry Leibowitz, Dave Yount helped mightily with statistics. Miki patiently encouraged. Mani supported and provided extensive critical review.